\documentclass[runningheads]{llncs}

\makeatletter
\newcommand{\@chapapp}{\relax}%
\makeatother
\usepackage[T1]{fontenc}
\usepackage{graphicx}
\usepackage{amsfonts}
\usepackage{amsmath}
\usepackage{hyperref}
\usepackage[title,header]{appendix}
\usepackage{changepage}
\usepackage{stmaryrd}
\usepackage{tikz}
\usetikzlibrary{shapes.symbols}
\begin{document}
\title{A 10-bit S-box generated by Feistel construction from cellular automata\thanks{This work has been supported by a government grant managed by the Agence Nationale de la Recherche under the Investissement d'avenir program, reference ANR-17-EURE-004}}
\titlerunning{10-bit S-box generated by Feistel construction from CA}
%
\author{Thomas Prévost\inst{1}\orcidID{0009-0000-2224-8574} \and
Bruno Martin\inst{1}\orcidID{0000-0002-0048-5197}}
\authorrunning{T. Prévost et al.}
%
\institute{Université Côte d'Azur, CNRS, I3S, France \\
\email{\{thomas.prevost,bruno.martin\}@univ-cotedazur.fr}}

%
\maketitle              
\begin{abstract}
We propose a new 10-bit S-box generated from a Feistel construction. The subpermutations are generated by a 5-cell cellular automaton based on a unique well-chosen rule and bijective affine transformations. In particular, the cellular automaton rule is chosen based on empirical tests of its ability to generate good pseudorandom output on a ring cellular automaton. Similarly, Feistel's network layout is based on empirical data regarding the quality of the output S-box.

We perform cryptanalysis of the generated 10-bit S-box, and we find security properties comparable to or sometimes even better than those of the standard AES S-box. We believe that our S-box could be used to replace the 5-bit substitution of ciphers like ASCON.

\end{abstract}
%
%

\section{Introduction}

Cryptography plays a central role in telecommunications, with symmetric encryption enabling secure communication between parties sharing a secret key. The ciphertext must appear random to outsiders lacking the key.

Symmetric ciphers fall into two categories: stream ciphers, which encrypt data on the fly, and block ciphers, which operate on fixed-size blocks. Block ciphers, such as AES, are more widely used. The core component of block ciphers is the substitution box (S-box), which introduces nonlinearity and must be carefully designed to resist cryptanalytic attacks such as linear~\cite{biryukov2011linear}, differential~\cite{biham2012differential}, and boomerang~\cite{wagner1999boomerang}. As S-boxes are permutations over $\llbracket0, 2^n\!-\!1\rrbracket$, the number of possible $n$-bit S-boxes grows factorially, making exhaustive analysis infeasible (e.g., $2^8! \approx 8.58 \cdot 10^{506}$).

While many constructions use algebraic techniques over finite fields, we propose a combinatorial method based on a Feistel network of depth 11. Our design combines three affine layers with eight layers of permutations generated by a 1-dimensional uniform binary cellular automaton, using a Boolean function as a pseudo-random permutation.

We begin with background on binary uniform cellular automata (CA), Boolean functions (Section~\ref{1dim-bcell_aut}), then review Feistel constructions and the Luby-Rackoff theorem (Section~\ref{luby_rackoff_sect}). Section~\ref{generation_lr} details our 10-bit S-box generation method, followed by its cryptanalysis (Section~\ref{cryptanalysis}). We conclude with applications and potential extensions.

\subsubsection{Related work}

Many research papers propose new methods for generating S-boxes. Most focus on 8-bit S-boxes, although some offer smaller S-boxes.

The generation of $2n$-bit S-boxes from $n$-bit subpermutations has already been considered in the literature, either by Feistel or MISTY constructions~\cite{li2014constructing,canteaut2015construction}.

\cite{nyberg1993differentially} proposed a method to generate S-Boxes of arbitrary size allowing to maximize their degree, their nonlinearity and to minimize their differential uniformity.

The use of CAs for cryptography is not recent~\cite{daemen1993framework}. Gutowitz proposed to use CAs for a block cipher~\cite{gutowitz1993cryptography}. Several papers have already been proposed to construct S-boxes or hash-functions from CA~\cite{mariot2019cellular}. However, to our knowledge, no one has yet designed an S-box from a CA based Luby-Rackoff construction.


\section{Definitions and notation} \label{1dim-bcell_aut}

\subsection{Uniform binary 1-dimensional cellular automaton}

Let us recall the definition of a uniform binary 1-dimensional cellular automaton.
\begin{definition}[Cellular automaton]
A cellular automaton is a triple $(\mathbb{F}_2, \delta, N)$ where:
\begin{itemize}
    \item $\mathbb{F}_2$ is the set of states.
    \item $\delta$ is the local transition function (or rule) $\mathbb{F}_2^n \longrightarrow \mathbb{F}_2$ with \emph{arity} $n$. 
    \item $N \subseteq \mathbb{Z}$ is the finite neighborhood, $card(N) = a$ being the size of the CA.
\end{itemize}
    
\end{definition}

\subsection{Boolean functions} \label{boolean_functions}

A Boolean function with $n$ variables  takes $n$ Boolean values as input and returns a single Boolean value as output. The local transition functions of CAs can be viewed as Boolean functions. There are $2^{2^n}$ $n$-variable Boolean functions.


Boolean functions are characterized by their truth table, which lists the outputs corresponding to the inputs, unique in the set of functions with a given number of variables. A convenient way to represent Boolean functions is given by their Algebraic Normal Form:

\begin{definition}[Algebraic Normal Form] \label{anf_def}
    Any $n$-variable Boolean function $f$ can be expressed by a unique binary polynomial,
     called Algebraic Normal Form (ANF): 
     $f(x) = \bigoplus_{u\in \mathbb{F}^n_2}a_u(\prod^n_{i=1}x_i^{u_i}), a_u\in\mathbb{F}_2, u_i$ i-th projection of u, $x_i$ being the i-th bit of input $x$.
\end{definition}

\begin{definition}[Algebraic degree]
    The algebraic degree of a function $f$ counts the number of variable in the largest monomial
     $x_1^{u_1} ... x_n^{u_n}$ of its ANF.
\end{definition}

A function $f$ is \emph{nonlinear} if and only if its degree is at least 2.

\begin{definition}[Hamming weight]
    The Hamming weight of a Boolean function $f$, written $w_h(f)$, is the number of
     $x \in \mathbb{F}_2^n$ such that $f(x)=1$.
\end{definition}

\begin{definition}[Balancedness] \label{balanced_def}
    A $n$-variable Boolean function $f$ is balanced if and only if $w_h(f) = 2^{n-1}$ (it returns as many ones as zeroes).
\end{definition}

\begin{definition}[Correlation-immunity] \label{correlation_immune_def}
    An $n$-variable Boolean function $f$ is k-correlation immune, $1 \le k \le n$, if and only if
     for any binary random input $x = x_1, ..., x_n$, $f(x)$ is statistically
     independent from any subset of size $k$ of $x$.
\end{definition}

The Walsh-Hadamard transform~\cite{carlet12005supports} of a Boolean function $f$ is an essential tool for analyzing the statistical properties of a Boolean function. It is defined by: $\hat{f}(\omega) = \sum_{x=0}^{2^n-1}(-1)^{f(x) \oplus x \cdot \omega}$, where $x \cdot \omega$ denotes the dot product of the two binary vectors.

\begin{theorem} \label{xiao_massey_theorem}
    A $n$-variable Boolean function $f$ is k-order correlation immune, $1 \le k \le n$ if and only if for every $\omega \in \mathbb{F}_2^n$ such that $1 \le w_h(\omega) \le k$, $\hat{f}(\omega) = 0$.
\end{theorem}

Xiao and Massey proved theorem~\ref{xiao_massey_theorem} in~\cite{xiao1988spectral}. A Boolean function that is both balanced and correlation immune at order $k$ is said to be \emph{resilient at order $k$}.

\begin{definition}[Strict avalanche criterion] \label{sac_def}
    A $n$-variable Boolean function $f$ satisfies the Strict Avalanche Criterion (SAC) iff
     $\forall i \in \llbracket1, n\rrbracket$, flipping the i-th bit of the input x results in the output
     $f(x)$ being changed for exactly half of the inputs x.
\end{definition}

The strict avalanche criterion is particularly interesting in the cryptographic
 context since it makes it difficult to infer input from output.

\section{Feistel constructions} \label{luby_rackoff_sect}

The Feistel construction~\cite{feistel1973cryptography} is a method for constructing secure pseudo-random bijective permutations from pseudo-random functions. The Feistel network, from a certain depth, guarantees the computational indistinguishability of its pseudo-random permutation from a random permutation. 

\begin{definition}
    A function $f: \mathbb{F}_2^n \longrightarrow \mathbb{F}_2^n$ is said to be \emph{pseudo-random} (PRF) if its output is computationally difficult to distinguish from a random output.
\end{definition}

\begin{definition}
    A pseudo-random function $f: \mathbb{F}_2^n \longrightarrow \mathbb{F}_2^n$ is called \emph{pseudo-random permutation} (PRP) if and only if it is bijective.
\end{definition}

\begin{figure}
\centering
\scalebox{.5}{\begin{tikzpicture}
\draw[draw=black] (-2.5,0) rectangle ++(5,1);
\node[text centered] at (0,0.5) {\textbf{Initial block}};
\draw [-to, thick](-1.75,0) -- (-1.75,-1);
\draw [-to, thick](1.75,0) -- (1.75,-1);

\draw[draw=black] (-3,-2) rectangle ++(2.5,1);
\node[text centered] at (-1.75,-1.5) {\textbf{L0}};
\draw[draw=black] (0.5,-2) rectangle ++(2.5,1);
\node[text centered] at (1.75,-1.5) {\textbf{R0}};

\draw[black] (1.75,-4) circle (1);
\node[text centered] at (1.75,-4) {$f_1$};
\draw [-to, thick](1.75,-2) -- (1.75,-3);
\draw [-to, thick](1.75,-5) -- (1.75,-5.5);
\node[text centered] at (1.75,-5.7) {\LARGE $\oplus$};
\draw [-to, thick](1.75,-5.9) -- (1.75,-6.5);
\draw [-to, thick](-1.75,-2) -- (1.60,-5.6);
\draw [-to, thick](1.75,-2) -- (-1.75,-6.5);

\draw[draw=black] (-3,-7.5) rectangle ++(2.5,1);
\node[text centered] at (-1.75,-7) {\textbf{L1}};
\draw[draw=black] (0.5,-7.5) rectangle ++(2.5,1);
\node[text centered] at (1.75,-7) {\textbf{R1}};

%

\draw [-to, thick](-1.75,-7.5) -- (-1.75,-8.5);
\draw [-to, thick](1.75,-7.5) -- (1.75,-8.5);
\draw[draw=black] (-2.5,-9.5) rectangle ++(5,1);
\node[text centered] at (0,-9) {\textbf{Final block}};
\end{tikzpicture}}
\caption{Example of Feistel construction of depth 1. $f_1$ is a pseudo-random permutations} \label{simple_lr}
\end{figure}
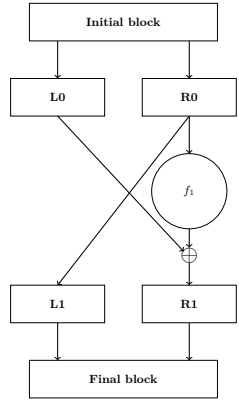

As shown in Fig.~\ref{simple_lr}, the Feistel construction creates a block permutation function of size $n$. It is made up of a stack of layers, each composed of PRP $f_i$ of input and output size $\frac{n}{2}$. We call \emph{depth} the number of sub-permutations $f_i$.

Luby and Rackoff~\cite{luby1988construct} proved that the output of the LR function is computationally indistinguishable from a random output as long as the depth of the network is at least 4, even for an adversary who knows the input. As shown by~\cite{patarin2003luby}, a Feistel construction with a depth of at least 7 returns an output that is computationally indistinguishable from a random output for an adversary able to choose the input value (Chosen-Plaintext-Attack, CPA), that is to say, no probabilistic algorithm is capable of making the distinction in polynomial time.

\section{Our 10-bit S-box from a CA based Feistel construction} \label{generation_lr}

\subsection{Architecture of the Feistel construction}

The permutation function generated by the Feistel construction allows to construct an S-box. We pass the $2^{10} = 1024$ possible inputs to the function, the output of which gives us the S-box. The latter must validate several security requirements, explained in section~\ref{cryptanalysis}. Another important property to respect is bijectivity, which allows to invert the S-box and permits decryption.

In our network, we use a 5-cell CA as a PRP $f_i$, its output is evaluated after a single time step on the input. The automaton has only one local transition function with 5 variables, which we detail in section~\ref{local_rule}.

However, a uniform CA will fail to return chaotic output for some particular inputs that are regular. If we used only this type of CA for intermediate permutation functions $f_i$, some inputs would still return a predictable result, for example $S(0)$ or $S(1023)$ which would return 0 or 1023 (0b1111111111).

Fortunately,~\cite{naor1997construction} tells us that it is possible to replace certain pseudo-random permutations by pair-wise independent permutations, i.e. permutations whose output is ``almost'' uniformly distributed for any two given inputs. An affine function $f_{a,b}(x) = a \cdot x + b$ satisfies these requirements. So that the permutation is bijective, we chose $a$ and $b$ prime.

The Luby-Rackoff construction we chose to generate our 10 bits S-box consists of the
 following eleven layers:
\begin{enumerate}
    \item A first layer uses the affine function $f_{5,3}(x) = 5.x + 3 \mod 2^{10}$.
    \item Next comes 4 layers using the 5-bit CA of section~\ref{local_rule} as PRP.
    \item The next layer uses the affine function $f_{7,11}(x) = 7.x + 11 \mod 2^{10}$.
    \item Next we have 3 layers of CA.
    \item We have another affine layer, $f_{13,17}(x) = 13.x + 17 \mod 2^{10}$.
    \item Finally a last layer reuses the 5-bit-cellular automaton.
\end{enumerate}

\subsection{Construction of the local PRP with a CA} \label{local_rule}

\subsubsection{Construction of the cellular automaton}

We are looking for a CA which permutes its input value. For this, we build a CA in a ring of 5 cells. To perform the permutation, we assign to the cells of the ring the value to be permuted, then we return the value of the CA after a single time step. We chose this construction because there are no 4-variable Boolean functions with convenient cryptographic properties and which allows to create a bijective CA.
\subsubsection{Basic properties of local transition rule}

There are $2^{2^5} = 4.294.967.296$ 5-bit local transition functions. The papers~\cite{wolfram1983statistical,formenti2014advances} fortunately give us some ideas for selecting the Boolean functions most likely to introduce ``chaos'' into the output of the cellular automaton.

Let us start by keeping only balanced Boolean functions (see def.~\ref{balanced_def}). There are then $\binom{32}{16} = 601.080.390$ functions left. We then eliminate the functions which are not first-order correlation immune (see def.~\ref{correlation_immune_def}), to keep only 807.980 rules. We next eliminate linear functions, to keep 807.928 functions. We are satisfied with the nonlinearity property here:~\cite{poinsot2006boolean} proves that there cannot exist bent functions (maximally non-linear) with an odd number of variables. Finally we eliminate all functions that do not respect the Strict Avalanche Criterion (SAC) (see def.~\ref{sac_def}), to keep 7.080 local rules.

\subsubsection{Selection over NIST FIPS 140-2 randomness test} \label{nist}

We try to create a pseudo-random generator from a uniform CA, then we only keep rules which allows to create ``good'' pseudo-random bit generators. We start by creating a ring CA, of size 1024 bits. Indeed,~\cite{preneel1993analysis} informs us that the ring must have more than 1000 cells to produce a secure pseudo-random generator.

For the seed value, we fill the ring with 1024 truly random bits. The seed and CA random generator are available at: \url{https://github.com/thomasarmel/cellular_automata_prng}.
 
Next, we test each of the 7.080 local rules as follows: at each time step, we update all cells in the ring with the rule under test. We then extract bit 512 from the ring. We iterate the CA and repeat the operation for as many bits as we wish to extract.

For each of the 7.080 rules, we evaluate the pseudo-random bit generator with the NIST FIPS 140-2 test~\cite{pub1994security}. The generator must pass all tests (``Monobit'', ``Poker'', ``Runs'', ``Long run'' and ``Continuous run'') out of 100.000 bits generated, which is equivalent to passing each test 39 times. There then remain 53 rules which allow the ring cellular automaton to validate the NIST FIPS 140-2 pseudo-random generation test.


Finally, we must ensure that the 5-bit ring cellular automaton of the Luby-Rackoff construction is bijective. To do this, we eliminate all the local rules which do not allow us to create a bijective CA.

Finally, only one rule remains, whose truth table numbering is $1.438.886.595$ (in decimal), or $x_0 \cdot x_3 + x_1 \cdot x_3 + x_2 \cdot x_3 + x_3 \cdot x_4 + x_1 + x_2 + x_3 + 1$.

The source code to generate the S-box can be found at the following address: \url{https://github.com/thomasarmel/luby_rackoff_sbox_finder}.

\section{Cryptanalysis} \label{cryptanalysis}

Here we propose the cryptanalysis of the specific S-box that we generated from the construction described above. In order to have a more systemic analysis of the security of S-boxes generated from Feistel networks, refer to~\cite{canteaut2015construction}.

Our S-box has a minimum algebraic degree of \textbf{8} and a maximum degree of 9. In comparison, the minimum and maximum degree of the AES S-box is 7 (having a larger S-box gives us an advantage). Our S-box's algebraic complexity is \textbf{1023}, which is the maximum possible. The algebraic complexity of the AES S-box is 255, which is also the highest possible value. (The algebraic complexity of an S-box defines its ability to resist interpolation attacks~\cite{jakobsen2001attacks}).

A strong nonlinearity~\cite{carlet2007nonlinearities} allows the S-box to resist linear cryptanalysis. It is defined as the minimum nonlinearity of each of the component functions. The nonlinearity of our S-box is \textbf{434}. For both AES S-box and ours, it is not possible to express the value of one of the output bits as a function of a linear combination of the input bits with a probability $\ge 60\%$ (56.25\% for AES S-box and 57.62\% for our S-box). 

The notion of \emph{Strict avalanche criterion} (SAC) (see def.~\ref{sac_def}) for the design of S-boxes was first introduced in 1985 by~\cite{webster1985design}. To satisfy the SAC, half of the output bits must be modified when a single input bit is modified. For an S-box, the bits of the SAC dependency matrix must be close to the ideal value of 0.5. Our S-Box has an average SAC of \textbf{0.50}, with a minimum of \textbf{0.44} and a maximum of \textbf{0.57}. For comparison, AES's S-Box has an average SAC of 0.50, with a minimum of 0.45 and a maximum of 0.56. So our average value is good, and the extreme values are almost as good as those of the AES S-box.

The \emph{Bit Independence Criterion Parameter} measures how inverting an input bit changes the output bits independently. The BIC parameter of our S-box is \textbf{0.124}. For comparison, the one of the AES S-box is 0.134. Our BIC parameter is therefore better than the one of the AES S-box.

The \emph{Linear Approximation Probability} (LAP) gives us an indication of how resistant our S-box is to linear cryptanalysis~\cite{biryukov2011linear}. The Linear Approximation Probability of our S-box is \textbf{9.28\%}, which is comparable or even better than other S-boxes proposed in the scientific literature~\cite{arshad2022novel}. However, the LAP is a little bit worse than that of the AES S-box, which is 6.25\%.

The \emph{Differential Approximation Probability} is determined by the XOR distribution between the input and output of an S-box. The lowest possible value guarantees the security of the S-box against differential cryptanalysis~\cite{biham2012differential}. The DAP of our S-box is \textbf{1.37\%}, in comparison the AES S-box DAP is 1.56\%.

The \emph{Differential Uniformity} of an S-box defines its proximity to perfect non-linearity~\cite{nyberg1992provable}. The Differential Uniformity of our 10-bit S-box is \textbf{14}. Let us divide this value by $\frac{2^{10}}{2^8}=4$ in order to compare with 8-bit S-box, we obtain 3.5. This is a better value than the Differential Uniformity of the AES S-box which is 4.

The \emph{Boomerang Uniformity} $\mathcal{BU}$ defines the resistance of an S-box to the boomerang attack~\cite{wagner1999boomerang}, which is an improvement of differential cryptanalysis. Our 10-bit S-box has a $\mathcal{BU}$ of \textbf{24}. Let's divide this value by 4 to compare it with 8-bit S-boxes. We find a value of 6, which equals the $\mathcal{BU}$ of the AES S-box.

\section*{Discussion}

Our 10-bit S-box demonstrates security properties comparable to the AES standard. Its quality may be further enhanced by increasing the Feistel network depth or adjusting affine function parameters, although we expect similar security behavior across such variations.

Larger S-boxes (e.g., 14- or 18-bit) could be constructed if $n$ is even and $\frac{n}{2}$ is odd, but selecting suitable Boolean functions with many variables (e.g., 7 or 9) becomes computationally intensive, as the number of $n$-bit Boolean functions grows exponentially (e.g., $1.34 \cdot 10^{154}$ for $n = 9$).

Our S-box could be integrated into sponge-based ciphers like ASCON~\cite{dobraunig2021ascon}, potentially improving security at the cost of performance. For instance, our Rust implementation (\url{https://github.com/thomasarmel/sponges/blob/sbox_10/ascon/src/lib.rs#L100}) shows a 10–15× slowdown on an Intel Core i7-13700H CPU, with added complexity in achieving constant-time execution.

\section*{Conclusion}

We have proposed a new 10-bit S-box from a Feistel construction based on uniform CA permutations, and carried out its cryptanalysis. In particular, we evaluated its robustness against linear, differential or boomerang attacks. We shown that our S-box has comparable, or even better security than those of the AES S-box. Our method can be extended for the construction of $n$-bit S-box, given that $n$ is even and $\frac{n}{2}$ is odd.


%
%
%
%
\bibliographystyle{splncs04}
\bibliography{sbox_from_lr}

\begin{thebibliography}{10}
\providecommand{\url}[1]{\texttt{#1}}
\providecommand{\urlprefix}{URL }
\providecommand{\doi}[1]{https://doi.org/#1}

\bibitem{arshad2022novel}
Arshad, B., Siddiqui, N., Hussain, Z., Ehatisham-Ul-Haq, M.: A novel scheme for
  designing secure substitution boxes ({S}-boxes) based on {M}öbius group and
  finite field. Wireless Personal Communications  (2022)

\bibitem{biham2012differential}
Biham, E., Shamir, A.: Differential cryptanalysis of the data encryption
  standard. Springer Science \& Business Media (2012)

\bibitem{biryukov2011linear}
Biryukov, A., De~Canniere, C.: Linear cryptanalysis for block ciphers.
  Encyclopedia of cryptography and security  (2011)

\bibitem{canteaut2015construction}
Canteaut, A., Duval, S., Leurent, G.: Construction of lightweight {S}-boxes
  using {Feistel} and {MISTY} structures. In: Selected areas in crypto.
  Springer (2015)

\bibitem{carlet2007nonlinearities}
Carlet, C., Ding, C.: Nonlinearities of {S}-boxes. Finite fields and their
  applications  (2007)

\bibitem{carlet12005supports}
Carlet, C., Mesnager, S.: On the supports of the {W}alsh transforms of
  {B}oolean functions. BFCA'05  (2005)

\bibitem{daemen1993framework}
Daemen, J., Govaerts, R., Vandewalle, J.: A framework for the design of one-way
  hash functions including cryptanalysis of {D}amg{\aa}rd's one-way function
  based on a cellular automaton. In: ASIACRYPT. Springer (1993)

\bibitem{dobraunig2021ascon}
Dobraunig, C., Eichlseder, M., Mendel, F., Schl{\"a}ffer, M.: Ascon v1. 2:
  Lightweight authenticated encryption and hashing. Journal of Cryptology
  (2021)

\bibitem{feistel1973cryptography}
Feistel, H.: Cryptography and computer privacy. Scientific american  (1973)

\bibitem{formenti2014advances}
Formenti, E., Imai, K., Martin, B., Yun{\`e}s, J.: Advances on random sequence
  generation by uniform cellular automata. Computing with new resources  (2014)

\bibitem{gutowitz1993cryptography}
Gutowitz, H.: Cryptography with dynamical systems. In: Cellular Automata and
  Cooperative Systems. Springer (1993)

\bibitem{jakobsen2001attacks}
Jakobsen, T., Knudsen, L.R.: Attacks on block ciphers of low algebraic degree.
  Journal of Cryptology  (2001)

\bibitem{li2014constructing}
Li, Y., Wang, M.: Constructing {S}-boxes for lightweight cryptography with
  {Feistel} structure. In: Int. Workshop on Crypto. Hardware and Embedded
  Systems. Springer (2014)

\bibitem{luby1988construct}
Luby, M., Rackoff, C.: How to construct pseudorandom permutations from
  pseudorandom functions. SIAM Journal on Computing  (1988)

\bibitem{mariot2019cellular}
Mariot, L., Picek, S., Leporati, A., Jakobovic, D.: Cellular automata based
  {S}-boxes. Cryptography and Communications  (2019)

\bibitem{naor1997construction}
Naor, M., Reingold, O.: On the construction of pseudo-random permutations:
  {L}uby-{R}ackoff revisited. In: ACM symposium on Theory of computing (1997)

\bibitem{nyberg1992provable}
Nyberg, K., Knudsen, L.R.: Provable security against differential
  cryptanalysis. In: Annual international cryptology conference. Springer
  (1992)

\bibitem{nyberg1993differentially}
Nyberg, K.: Differentially uniform mappings for cryptography. In: Workshop on
  the Theory and Application of Cryptographic Techniques. Springer (1993)

\bibitem{patarin2003luby}
Patarin, J.: {L}uby-{R}ackoff: 7 rounds are enough for $2 n (1 - \varepsilon$)
  security. In: CRYPTO. Springer (2003)

\bibitem{poinsot2006boolean}
Poinsot, L.: {B}oolean bent functions in impossible cases: odd and plane
  dimensions. International Journal of Computer Science and Network Security
  (2006)

\bibitem{preneel1993analysis}
Preneel, B.: Analysis and design of crypto. hash functions. Ph.D. thesis (1993)

\bibitem{pub1994security}
Pub, F.: Security requirements for cryptographic modules. FIPS PUB  (1994)

\bibitem{wagner1999boomerang}
Wagner, D.: The boomerang attack. In: FSE. Springer (1999)

\bibitem{webster1985design}
Webster, A.F., Tavares, S.E.: On the design of {S}-{B}oxes. In: Conference on
  the theory and application of cryptographic techniques. Springer (1985)

\bibitem{wolfram1983statistical}
Wolfram, S.: Statistical mechanics of ca. Reviews of modern physics  (1983)

\bibitem{xiao1988spectral}
Xiao, G.Z., Massey, J.L.: A spectral characterization of correlation-immune
  combining functions. IEEE Transactions on information theory  (1988)

\end{thebibliography}

\end{document}